\documentclass[12pt,reqno]{article}
\usepackage{amsmath}
\usepackage{amsfonts}
\usepackage{amssymb}
\usepackage{latexsym}
\usepackage[dvips]{graphicx}
\usepackage{multirow}
\usepackage{color}
\usepackage{cite}
\usepackage{subfigure}
\usepackage{dsfont}
\usepackage{booktabs,multirow}

\def\OM{\Omega_{\rm M}}
\def\OL{\Omega_{\Lambda}}

\def\be{\begin{equation}}
\def\ee{\end{equation}}     

\begin{document}
%
%
\begin{center}
{\LARGE\bfseries{  
Dynamical System Analysis of Cosmologies\\
 with Running Cosmological Constant from\\[3mm] 
Quantum Einstein Gravity} }
\\[10mm]
Alfio Bonanno$^{1,2}$  and Sante Carloni$^3$ \\[3mm]
{\small\slshape
$^1$INAF, Osservatorio Astrofisico di Catania, Via S.Sofia 78, 95123 Catania Italy \\
$^2$INFN, Sezione di Catania, Via S.Sofia 72, 95123 Catania, Italy \\
$^3$ESA-Advanced Concept Team, European Space Research Technology Center (ESTEC)\\
Keplerlaan 1, Postbus 299, 2200 AG Noordwijk The Netherlands.}

\end{center}

\vspace{8mm}
\vspace{8mm}
\hrule\bigskip
\centerline{\bfseries Abstract} \medskip \noindent
We discuss a mechanism that induces a time-dependent  vacuum energy
on cosmological scales.  It is based on the instability induced renormalization 
triggered by  the  low energy quantum fluctuations in a Universe with a positive cosmological constant.
We  employ the dynamical systems approach to study the qualitative behavior of 
Friedmann-Robertson-Walker cosmologies where the cosmological constant is dynamically 
evolving according with this nonperturbative scaling at low energies.
It will be shown that it is possible
to realize a ``two regimes'' dark energy phases, where an unstable early phase
of power-law evolution of the scale factor is followed by an accelerated expansion
era at late times.
\noindent

\bigskip
\hrule\bigskip

\newpage

\section{Introduction}
The idea that Quantum Gravity effects can be important at astrophysical and cosmological distances 
has recently attracted much attention.  In particular the framework of
Exact Renormalization Group (ERG) approach for quantum gravity
\cite{1998PhRvD..57..971R} has opened the possibility of investigating both the ultraviolet
(UV) and the infrared (IR) sector of gravity in a systematic manner. 

The essential ingredient of this tool is the Effective Average Action
$\Gamma_k[g_{\mu\nu}]$, a Wilsonian coarse grained free energy dependent on
an infrared momentum scale $k$ which defines an effective
field theory appropriate for the scale $k$. By construction, when evaluated at ${\it tree}$ level,
$\Gamma_k$ correctly describes all gravitational phenomena, {\it including all loop effects},
if the typical momentum involved are of the order of $k$.
When applied to the Einstein-Hilbert action the ERG yields
renormalization group flow equations
\cite{1998CQGra..15.3449D} 
which have made possible detailed investigations of the scaling behavior of NewtonÕs constant 
at high energies
\cite{1999PThPh.102..181S,2002PhRvD..65b5013L,2002PhRvD..66b5026L,2002CQGra..19..483L,
2002PhRvD..65f5016R,
2002PhRvD..66l5001R,
2004PhRvL..92t1301L,
2005JHEP...02..035B,
2006PhRvL..97v1301C,
2009GReGr..41..983R,
2009PhRvD..79j5005R,
2009arXiv0904.2510M,
2009arXiv0905.4220M}.
The  scenario emerging from these studies,  
first demonstrated by Weinberg \cite{wein2} in $d=2+\epsilon$ dimensions, suggests that the theory could be
consistently defined  in $d=4$ at a nontrivial UV fixed point where the 
dimensionless Newton constant,
$g(k)=G(k) k^2$, 
does not vanish in the $k\rightarrow\infty$ limit, i.e. $g(k\rightarrow\infty)=g^\ast$. 
As a consequence the dimensionful Newton  constant $G(k)$
is {\em antiscreened} at high energies, very much as one would expect based on
the intuitive picture that the larger is the cloud of virtual particles, the
greater is the effective mass seen by a distant observer
\cite{1993hep.th....4146P}.

Recent works have included matter fields \cite{2003PhRvD..67h1503P} and have
also  considered a growing number of purely gravitational operators in the
action. In particular, truncations involving quadratic terms in the curvature have been
considered in \cite{2006PhRvL..97v1301C,niedermaier_asymptotic_2006,2009AIPC.1196...44B,2010NuPhB.824..168B}, 
while higher powers of the Ricci scalar have been studied in 
\cite{2009AnPhy.324..414C,2008IJMPA..23..143C}. In all the investigations the   
UV critical surface has turned out to be finite dimensional ($d_{\rm UV}=3$), implying that the theory is
nonperturbatively renormalizable. At the UV fixed point 
the theory has a behaviour very similar to QCD, being weakly coupled at high
energies  but  
the running of the dimensionful Newton constant $G(k)=g(k)/k^2$  in  the
deep ultraviolet region is  a power-law, at variance with the logarithmic scaling of QCD.

A weakly coupled gravity at high energies is expected to generate
important consequences in several astrophysical and cosmological contexts
\cite{2009arXiv0911.2727B}, and in fact 
the RG flow of the effective average action, obtained by
different truncations of theory space, has been the basis of various investigations of 
``RG improved" black hole spacetimes, 
\cite{1999PhRvD..60h4011B,2000PhRvD..62d3008B,2006PhRvD..73h3005B} 
and Early Universe models
\cite{2002PhRvD..65d3508B,2003GReGr..35.1899B,2004CQGra..21.5005B,2005IJMPA..20.2358B,2006CQGra..23.3103B,2006rdgp.conf..461B}.

However, the behavior of the theory is more complicated 
at low energies, corresponding at cosmological 
scales, because the $\boldsymbol{\beta}$--functions
of any local operator of the type $\sqrt{g}R$, $\sqrt{g}R^2$, \ldots,
$\sqrt{g}R^n$ are singular in the IR due to the presence of a pole at
$\Lambda(k)/k^2=1/2$ being $\Lambda$ the cosmological constant. The presence of this pole is signaling that the
Einstein-Hilbert truncation is no longer a consistent approximation to the full flow equation, and most
probably a new set of IR-relevant operators is emerging in the $k\rightarrow 0$
limit. This singular behavior  in fact appears in the ERG for  nearly all cutoff
threshold functions in the Einstein-Hilbert truncation, and it is caused
by the presence of negative eigenvalues in the spectrum of the fluctuations determinant
of the gravitational degrees of freedom.
As discussed in \cite{2004JCAP...12..001R} the dynamical origin of these strong IR
effect is due to the ``instability driven renormalization'', a phenomenon well known from many other
physical systems
\cite{1999PhLB..445..351A,2000PhRvD..62l5021L,2004NuPhB.693...36B}.
We shall see that the low energy domain 
of the theory is regulated by an IR
fixed point which drives the cosmological constant to zero at very late times
$\Lambda(t \rightarrow \infty)=0$. 
This new-type of ``decaying $\Lambda$
cosmologies'' is therefore quite different from previous models where the
time-dependent cosmological constant encodes the effect of the matter creation
process \cite{1996PhRvD..54.2571L}.

Astrophysical consequences of the possible presence of an IR fixed point for quantum
gravity  appeared in
\cite{2002PhLB..527....9B,2004IJMPD..13..107B,2004JCAP...01..001B,2006CQGra..23.3103B,2004ForPh..52..650R} 
where it was shown that a solution of the ``cosmic coincidence problem'' 
arises naturally without the introduction of a  ``dark energy field''. 
In particular in the fixed point regime the
vacuum energy density $\rho_\Lambda\equiv\Lambda/8\pi G$
is automatically adjusted so as to equal the matter energy density,
{\it i.e.} $\OL=\OM=1/2$, and the deceleration parameter approaches $q
= -1/4$. Moreover, an analysis of the high-redshift SNe Ia data leads to the conclusion that
this {\em infrared fixed point cosmology} is in good  agreement with
the observations \cite{2004JCAP...01..001B}.
Recent works have also considered the possibility that the ``basin of
attraction'' of the IR fixed point can act already at galactic scale, thus
providing an explanation for the galaxy rotation curve without dark matter 
\cite{2004JCAP...12..001R,2004PhRvD..69j4022R,2004PhRvD..70l4028R,2007CQGra..24.6255E},
although a detailed analysis based on available experimental data is still missing.
Cosmologies discussing the complete evolution from the early Universe to the present time
have been considered in \cite{2005JCAP...09..012R} and in
\cite{2007JCAP...08..024B}. 
In particular in \cite{2007JCAP...08..024B} it has been shown
that the ``RG improved'' Einstein equations admit (power-law or exponential) inflationary solutions
and that the running of the cosmological constant can account
for the entire entropy of the present universe in the massless sector 
(see also \cite{2008JPhCS.140a2008B} for a review).

The main purpose of this paper is to explore the idea that quantum effects
could dynamically drive the cosmological constant to zero at late times
so that $\Lambda(t\rightarrow\infty)=0$ as a result  
of an explicit dynamical mapping $k\rightarrow k(t)$ of the renormalization group trajectories  
generated by the unstable infrared modes of the gravitational sector.

{This idea will then be studied using the so-called 
Dynamical Systems Approach (DSA), a  
technique  already used in cosmology by 
Bogoyavlensky \cite{oleg}
and further 
developed by Collins, and  Ellis and Wainwright  
to analyze non trivial cosmologies (i.e. Bianchi models) in the context of pure General Relativity \cite{ellisbook}. 
Some work has also been done in the case of (minimally coupled) scalar fields in cosmology \cite{Copeland:2009be,Copeland:1997et} 
and, more recently, of scalar tensor theories of gravity \cite{Carloni:2007eu},  $f(R)$ theories of gravity 
\cite{Carloni:2004kp,Leach:2006br,Abdelwahab:2007jp,Carloni:2007eu,Carloni:2007br} 
and  Ho\v{r}ava-Lifschits gravity \cite{Carloni:2009jc}. 
Studying cosmologies using the DSA has the advantage of offering a relatively simple method to 
obtain particular exact solutions and to obtain a (qualitative) description of the global dynamics of these models.}

{In our specific context the DSA allows us to prove that}
the presence of  a singular behavior of the RG evolution for the
cosmological constant in the infrared put strong constraints on the possible RG
trajectories. In particular we will show that it is possible to realize a scenario 
where the Universe has a transition from an early unstable phase of 
power-law evolution of the type $a\propto t^\beta$ with $\beta \in ]0,1[$ to a
de-Sitter phase. 

The structure of the paper is the following. In Section 2 the basic mechanism
of the dynamical suppression of the cosmological constant  by the unstable low energy modes
is discussed. In Section 3 the dynamical system analysis of the
resulting cosmologies is presented. Section 4 is devoted to the conclusions. 

\section{Instability induced renormalization}
It is interesting to review in detail the main arguments suggesting that
the cosmological constant $\Lambda$  must have a nontrivial scaling at 
cosmological distances due to QG effects  
\cite{2004JCAP...12..001R}.
As already mentioned, the key
mechanism is the so called ``instability induced renormalization
\cite{1999PhLB..445..351A,2000PhRvD..62l5021L,2004NuPhB.693...36B}.
In order to illustrate this point let us look to  
$\mathcal{Z}_{2}$--symmetric real scalar field in a simple truncation:
\begin{equation}\label{sca}
{\Gamma}_{k} [\phi] = \int \!\! {d}^{4} x~
\Big \{
\frac{1}{2} \, \partial_{\mu} \phi \, \partial^{\mu} \phi
+ \frac{1}{2} \, m^{2} (k) \, \phi^{2}
+ \frac{1}{12} \, \lambda (k) \, \phi^{4}
\Big  \}.
\end{equation}
In a momentum representation we have
\begin{equation}
{\Gamma}^{(2)}_{k} =\frac{\delta^2 {\Gamma}_k}{\delta \phi^2} = p^{2} + m^{2} (k) +
\lambda (k) \, \phi^{2},
\end{equation}
so that ${\Gamma}^{(2)}_{k}$ is positive if $m^{2} >0$; but when $m^{2} <0$ it can
become negative for $\phi^{2}$ small enough. Of course, the negative eigenvalue for $\phi
=0$, for example, indicates that the fluctuations are unstable, and the
non-linear evolution of this instability is a ``condensation'' which 
shifts the field from the ``false vacuum'' to the true one, the
phenomenon that produces the instability induced renormalization. 
In particular, the
$\boldsymbol{\beta}$--functions,  obtained by $p$--integrals over
(powers of) the propagator
$
\Bigl[ p^{2} + m^{2} (k) + k^{2} \Bigr]^{-1}
$
are regular in the symmetric phase ($m^{2} >0$) 
but there is a pole at $p^{2} = - m (k)^{2} - k^{2}$ provided $k^{2}$ is small
enough in the broken phase ($m^{2} <0$).
For $k^{2} \searrow | m (k)^{2}|$ the
$\boldsymbol{\beta}$--functions become large and there the instability induced
renormalization occurs. In a reliable truncation, a physically realistic RG
trajectory in the spontaneously broken regime will not hit the singularity at $k^{2} = |
m (k)^{2}|$, but rather make $m (k)$ run in  such a way that
$| m (k)^{2}|$ is always smaller than $k^{2}$. This requires that
\be\label{qma}
- m (k)^{2} \propto k^{2}.
\ee
In order to avoid the singularity, a mass renormalization is necessary in
order to evolve a double-well shaped symmetry breaking classical
potential into an effective potential which is convex and has a flat
bottom, as it emerges from analytical and numerical calculations
\cite{2004NuPhB.693...36B}.
However, the truncation implied in (\ref{sca}) is not enough 
to describe the broken phase, because 
its RG trajectories terminate at a
finite scale $k_{\text{term}}$ with $k_{\text{term}}^{2} = | m
(k_{\text{term}})^{2}|$ at which the $\boldsymbol{\beta}$--functions
diverge. 
Instead, if one allows for an arbitrary running potential
$U_{k} (\phi)$, containing infinitely many couplings, all trajectories
can be continued to $k=0$, and for $k \rightarrow  0$ one finds indeed the
quadratic mass renormalization (\ref{qma}) as discussed in \cite{2004NuPhB.693...36B}.

In the case of gravity we can  consider a 
family of ``off-shell'', spherically symmetric backgrounds 
labeled by the radius of the sphere $\phi$,
in order to  disentangle the 
contributions from the two invariants $\int \! {d}^{4} x
\sqrt{g\,} \propto \phi^{4}$ and $\int \! {d}^{4} x \sqrt{g\,}
\, R \propto \phi^{2}$  to the Einstein-Hilbert flow.
It is then convenient to decompose the fluctuation $h_{\mu \nu}$ on the
sphere into irreducible  components \cite{2002PhRvD..65b5013L}
and to expand the irreducible pieces in terms of the corresponding
spherical harmonics. For $h_{\mu \nu}$ in the transverse--traceless
(TT) sector,  the operator $\Gamma^{(2)}_{k}$ equals, up
to a positive constant,
\be\label{pna}
-\nabla ^2 +  8 \, \phi^{-2} + k^{2} - 2 \, \Lambda (k),
\ee
where $\nabla^{2} \equiv g^{\mu \nu} \, \nabla_{\mu} \nabla_{\nu}$ is the
covariant Laplacian acting on TT tensors.  The spectrum of $- \nabla^{2}$, 
denoted $\{p^{2} \}$, is discrete and positive. Clearly (\ref{pna}) is a
positive operator if the cosmological constant is negative. In this
case there are only stable, bounded oscillations, leading to a mild
fluctuation induced renormalization. 
The situation is very different for $\Lambda > 0$ where, for $k^{2}$
sufficiently small, (\ref{pna}) has negative eigenvalues, i.\,e.\
unstable eigenmodes.


A consistent calculation including 
all the components of the metric fluctuation $h_{\mu\nu}$,  explicitly illustrates this scenario. 
Following \cite{2005JHEP...02..035B},  the   $\boldsymbol{\beta}$--functions
for the dimensionless Newton constant 
$g(k)\equiv k^{d-2}G(k)$ 
and the dimensionless cosmological constant
$\lambda(k)\equiv k^{-2}\Lambda(k)$
can be  introduced
\begin{eqnarray}
\label{bb}
&&\partial_t g = \beta_g(g,\lambda) \equiv [d-2+\eta_N]g,\\[2mm]
&&\partial_t \lambda = \beta_\lambda (g,\lambda),
\end{eqnarray}
where 
\begin{eqnarray}\label{16}
&&\beta_\lambda= -(2-\eta_N)\lambda 
+4(4\pi)^{1-\frac{d}{2}}\Big
[\frac{d(d+1)}{4}\;(1-2\lambda)^{\frac{d}{2}-n-1}-d\Big ]g
\frac{\Gamma(n+1-\frac{d}{2})}{\Gamma(n+1)},
\end{eqnarray}
being $\eta_N \equiv -G(k)^{-1}\partial_t G(k)$  the anomalous 
dimension. Its  explicit expression reads 
\begin{equation}\label{15}
\eta_N=8(4\pi)^{1-\frac{d}{2}}\Big [ \frac{d(7-5d)}{24}
(1-2\lambda)^{\frac{d}{2}-n-2}- \frac{d+6}{6}\Big ]g 
\frac{\Gamma(n+2-\frac{d}{2})}{\Gamma(n+1)},
\end{equation}
where $n>1$  is an integer related with the regulator and $d$ is the
dimension of the spacetime \cite{2005JHEP...02..035B}.

Clearly,  
the allowed part of the $g$-$\lambda$--plane ($\lambda < 1/2$) in (\ref{16}) and (\ref{15}),  
corresponds to the situation $k^{2} > 2 \,
\Lambda (k)$ where the singularity is avoided thanks to the large
regulator mass. When $k^{2}$ approaches $2 \, \Lambda (k)$ from above
the $\boldsymbol{\beta}$--functions become large and strong
renormalizations set in, driven by the modes that would go
unstable at $k^{2} = 2 \,\Lambda$.
In this respect the situation is completely analogous to the scalar
theory discussed above: its symmetric phase ($m^{2} >0$) corresponds
to gravity with $\Lambda <0$; in this case all fluctuation modes are
stable and only small renormalization effects occur. Conversely, in
the broken phase ($m^{2} <0$) and in gravity with $\Lambda >0$, there
are modes, which are unstable in absence of the IR regulator. They lead
to strong IR renormalization effects for $k^{2} \searrow | m (k)^{2}|$
and $k^{2} \searrow 2 \, \Lambda (k)$, respectively. 

We are thus led to the conclusion that the instability induced renormalization should  
occur also in this framework as $k\rightarrow 0$, so that to avoid the
singularity the cosmological constant must run proportional to $k^{2}$,
\be
\Lambda (k) = \lambda^{\rm{IR}}_{*} \, k^{2} + \mathrm{subleading \, \, terms}, 
\;\;\,\,\, k\rightarrow 0,
\label{ir}
\ee
with the constant $\lambda^{\mathrm{IR}}_{*} < 1/2$  
being {\it  an infrared fixed point} of the
$\lambda$--evolution. If the behavior (\ref{ir}) is actually
realized, the renormalized cosmological constant observed at very
large (cosmological) distances, $\Lambda (k \to 0)$, vanishes regardless of its bare
value. It is important to stress that recent investigations based on a 
conformal reduction of Einstein-Gravity have actually found a new IR fixed point
which could represent the counterpart, in the reduced theory,
of the physical IR fixed point present in the full theory
\cite{Manrique:2010p3084}.

As in the  case of the scalar field, the presence of an IR pole is
signaling that the Einstein-Hilbert truncation is no longer a consistent
approximation to the full flow equation near the IR singularity, and most
probably, a new set of IR-relevant operators is emerging ad $k\rightarrow 0$.
Although we do not have an explicit solution for the RG flow near
the unstable phase in the case of gravity\footnote{Even for the simple scalar theory the non-perturbative
investigation of the IR instability requires  
the use of special numerical techniques to correctly resolve the singularity,
see \cite{2004NuPhB.693...36B} for details.},
near the IR fixed point we can always write
\be
\frac{\Lambda(k)}{k^2}=\lambda^{\mathrm{IR}}_{*}+h_2 k^{2\theta}, \;\;\;\;
k\rightarrow 0,
\ee
being $\theta$ a critical exponent and $h_2$ a constant related with the eigenvalue of the stability 
matrix around the IR fixed point. Its precise value cannot be determined within the 
linearized theory, but we shall see that our analysis will not depend on the actual value 
of $h_2$.
For $\theta>0$ the IR fixed point is
attractive, while for $\theta<0$ is repulsive. In this latter case the 
IR fixed point is a {\it high-temperature fixed point} where
$\boldsymbol{\beta}$--functions are suppressed as $k\rightarrow 0$ and the flows
stops before reaching $\lambda^{\mathrm{IR}}_{*}$.

In order to  close the system we must map the RG flow onto the Universe dynamics
so that $k\rightarrow k(t)$.
Clearly, the Hubble scale $H(t)$ would be a natural choice for the infrared cutoff $k$ 
because  in
cosmology the Hubble length $1/H(t)$ measures the size of the
``Einstein elevator'' outside which curvature effects become appreciable
\cite{2007JCAP...08..024B,2005JCAP...09..012R},  and  
therefore $k \sim H (t)$.
For actual calculations we will set  
\begin{equation}\label{cui}
k(t)=\xi H (t),
\end{equation}
being $\xi$ a positive number which we expect should be of the order of unity. 
We are thus led to the following $\Lambda(H)$ dependence:
\begin{equation}\label{Lambda}
   \Lambda  =
   H^2\,{\xi }^2\,{{\lambda }_*} +H^{\alpha }\,{\xi }^{\alpha }\,{h_2} \equiv
   A\,H^2 + B\,H^{\alpha }, \qquad\alpha  = 2 +
   2\,\theta\; .
\end{equation}
Here $0<\alpha<2$ correspond with a repulsive IR fixed point, while $\alpha>2$
represents an attractive IR fixed point. The parameters $A=\xi^2 {\lambda }_*$
and $B={\xi }^{\alpha }\,{h_2}$
are introduced for  notation simplicity instead of $\xi$ and $h_2$ in the following discussion.
Remarkably, we shall see that our conclusions will be  independent on the actual value
of $B$, while the parameters $A$ and $\alpha$ will be the only free parameters of our theory. 

\section{Dynamical system analysis}
\subsection{Basic equations}
From the previous discussion it is then clear that in this framework the
cosmological constant $\Lambda$ is promoted to the status of dynamical
variable by Eq.(\ref{Lambda}), so that   $\Lambda=\Lambda(t)$.

Let us then specialize $g_{\mu\nu}$
to describe a generic Friedman-Robertson-Walker metric with scale factor $a(t)$,
where we take ${T_\mu}^\nu = {\rm diag}[\rho,-p,-p,-p]$ to be the energy momentum
tensor of an ideal fluid with equation of state $p=w\rho$ being $w\geq0$ a constant\footnote{ The dynamical systems analysis that will follow can be easily generalized to the case of non constant $w$ by, for example, adding a new variable corresponding to the pressure. However, this information does not add any further understanding of the relation between RG flow and dark energy and we will therefore  limit ourselves to the case of a single fluid with constant barotropic factor $w$. }.
Then the quantum gravity ``improved" Einstein equations reduce to
the following set of cosmological equations:
\begin{eqnarray}
   &&\frac{\kappa}{{a}^2} + {H}^2 = 
   \frac{8\,\pi \,G\,\rho }{3} +\frac{1}{3} \Lambda\label{RG cosm eq1},\\
&&\dot{H}+H^2=-\frac{4}{3} \pi  G \rho  (3 w +1) +\frac{1}{3}
\Lambda,\label{RG cosm eq2}\\ &&\dot{\rho} = -3\,\left( 1 + w  \right) \,H\,\rho  -
    \frac{\rho \,\dot{G}}{G} - \frac{\dot{\Lambda}}{8\,\pi \,G},\label{RG cosm eq3}
\end{eqnarray}
where $H=\dot a /a$ and $\kappa=-1,0,1$ with the usual meaning.
{It is difficult to determine 
the running of $G$ near the IRFP since the $\beta$-functions are singular near $k\rightarrow 0$, as discussed before.
However one expects that if an IRFP is also present in the running of $g$
so that $g(k\rightarrow 0)=g_\ast$, the dimensionful coupling constant 
$G(k)$ grows without bound as $k\rightarrow 0$, so that  $G(t\rightarrow \infty)=\infty$ at late times.}

In the following we will assume that energy momentum tensor of matter field is conserved
so that we are led to the following ``consistency condition'' 
\begin{equation}\label{G-anzat}
\dot{G}=-\frac{\dot{\Lambda}}{8\pi \rho}\;.
\end{equation}
and the dynamical evolution of $G$ is fixed by this request as a function of $\Lambda(t)$.
{We shall see that the late time behavior of $G$ obtained by the assumption (\ref{G-anzat}) is actually consistent
with the possibility that running of $g$ in the infrared is determined by an IRFP}.

{\
By using Eq.(\ref{Lambda}) in the Friedmann equation we find
\begin{equation}\label{useful fried}
  \frac{3\,\kappa}{{a}^2} + \left( 3 - A \right){H}^2 =BH^{\alpha}+8\,\pi \,G\,\rho,
\end{equation}
which resembles the standard GR one.
Assuming $H\neq0$ and $A\neq3$ , we  define dimensionless
variables
\begin{equation}\label{DynVar}
x = \frac{8\,\pi \,G\,\rho }{\left( 3 - A \right)
    \,{H}^2}\, ,\;\;\;\;\;\;\;
  y= \frac{B{H}^{\alpha-2} }{\left( 3 - A \right) }\, ,\;\;\;\;\;
  K= \frac{3\,\kappa }{\left( 3 - A \right)\,{H}^2 \,{a}^2}\;\;\;\;\;.
\end{equation}
which will be considered to be functions of  the logarithmic time $\mathcal{N}=\ln (a/a_0)$ where $a_0$ is the value of the scale factor at some reference time.
With some algebra  the
cosmological equations can be written as the first order
autonomous system
\begin{eqnarray}
  \nonumber x' &=& \frac{2A}{3} -
     \left[\frac{1}{3}(A+3)-\left(A - 3 \right)w   \right]x
     -\frac{1}{3}\left(A-3 \right) \,\left( 1 + 3\,w  \right) \,
     {x}^2 - \frac{1}{3}\left[\alpha (A-3) +2\,A \right]\,y\\
&&    + \frac{1}{3}\,\alpha\left( A-3 \right) \, \,{y}^2 + \frac{1}{6}\left(
    A-3\right)\,
       \left( 4 - \alpha \,\left( 1 + 3\,w  \right)  \right) \,x\,y ,\label{sistema non simpl1}\\
  y' &=&\left( 1 - \frac{A}{3} \right) \,\left( 2 - \alpha  \right) \,
    \left[ 1 + \frac{\left( 1 + 3\,w  \right) \, x}{2} - y \right] \,y,\label{sistema non simpl2}\\
    K' &=& K\,
    \left[
      \left( 1 - \frac{A}{3} \right) \,\left( 1 + 3\,w  \right) \,x -
      2\,\left( 1 - \frac{A}{3} \right) \,y -\frac{2}{3}A \right] ,\label{sistema non simpl3}
 \end{eqnarray}
with the constraint
\begin{equation}\label{constraint}
     1 + K - x - y = 0.
\end{equation}
In the above equations the prime stands for the derivative with respect to $\mathcal{N}$.

The (\ref{constraint}) allows to eliminate one of the equations of the (\ref{sistema non simpl1}-\ref{sistema non simpl3}) and obtain 
a two-dimensional phase space. We choose here to eliminate (\ref{sistema non simpl3})  to obtain:
\begin{eqnarray}\label{sistema simpl}
 \nonumber x' &=& \frac{2A}{3} -
     \left[\frac{1}{3}(A+3)-\left(A - 3 \right)\,w   \right] \,x
     -\frac{1}{3}\left(A-3 \right) \,\left( 1 + 3\,w  \right) \,
     {x}^2\\ && - \frac{1}{3}\left[\alpha (A-3) +2\,A \right]y
    + \frac{1}{3}\alpha\left( A-3 \right) {y}^2 + \frac{1}{6}\left(
    A-3\right)\,
       \left( 4 - \alpha \,\left( 1 + 3\,w  \right)  \right) \,x\,y ,\label{sistema simpl1}\\
  y' &=&\left( 1 - \frac{A}{3} \right) \,\left( 2 - \alpha  \right) \,
    \left[ 1 + \frac{\left( 1 + 3\,w  \right) \, x}{2} - y \right]\,y,\label{sistema simpl2}\\
  0 &=& 1 + K - x - y.\label{sistema simpl3}
\end{eqnarray}
Note that if $y=0$, the above system implies $y'=0$ and the $x$
axis is an invariant submanifold for the phase space. This means
that if the initial condition for the cosmological model is
$y\neq0$ a general orbit can approach $y=0$ only asymptotically.
As a consequence, there is no orbit that crosses the $x$ axis and
no global attractor can exist in the phase space.

\subsection{Finite Analysis}

Setting $x'=0,\, y'=0$ we obtain the three fixed points in Table \ref{tavola punti fissi}.

Two of these points ($\mathcal{A}$, $\mathcal{C}$) do not depend on the values of the parameters,
but one ($\mathcal{B}$) has the $x$ coordinate which is a
function of $A$ and the barotropic factor $w$. This fact
influences also the value of $K$ i.e. the sign of the spatial
curvature index $\kappa$. Merging will occur between $\mathcal{A}$ and
$\mathcal{B}$ for $A=\displaystyle{\frac{1+3w}{1+w}}$ and
between $\mathcal{B}$ and $\mathcal{C}$ for $A=0$. The first value
for $A$ represents a bifurcation for the dynamical system and we
will not treat it here, the second correspond to the case in which
$\Lambda$ does not contain any quadratic term as a function of $H$.

The solutions associated with the above fixed points can be
obtained integrating the equation
\begin{equation}\label{soluzioni}
  \dot{H}= \left( 1 - \frac{A}{3} \right) \,
   \left( -1 - \frac{1}{2}\left( 1 + 3\,w  \right) \,x + y\right){H}^2,
\end{equation}
and are listed in Table \ref{tavola punti fissi}. We can see that
$\mathcal{A}$ is a power law solution whose index resembles the
standard Friedmann solution, but it is modified by the parameter
$A$. It is easy to see that this point represents an expansion if
$A<3$ and a contraction if $A>3$. The point $\mathcal{B}$
corresponds to a Milne solution and $\mathcal{C}$ corresponds to a
de Sitter solution. 

The behavior of the energy density  and of the  gravitational ``constant'' can be only achieved once 
an assumption on $G$ has been made. Using  the  (\ref{G-anzat}) 
we find that the behavior of $G$ is in general given by combination of powers of $t$ which depend on $w$ and $\alpha$ 
as it emerges from the results depicted inTable \ref{tavola punti fissi}.
{However, note in particular that for $0<A<3$ for case ${\cal A}$ and for case ${\cal B}$ 
the evolution of $G$ implies a strongly coupled gravity 
at late times, as implied by the IRFP point model \cite{2002PhLB..527....9B}. }
Substitution in the field equations reveals that the point $\mathcal{A}$ represents 
a physical solution [i.e. satisfies the (\ref{RG cosm eq1}-\ref{RG cosm eq3})] 
only if $\kappa=0$, and  $B= 0$ while $A$ can take any value.

The solution for $\mathcal{B}$ is physical only for $B=0$,  the space curvature is given by 
 \begin{equation} \label{ttt}
\kappa=a_0^2 \left[\frac{A(1+ w)}{(1+3w)}-1\right]\,.
\end{equation}
so that this solution is not flat in general, (i.e. $\kappa\neq 0$). 
In the special case $\alpha=2$, $B$ is not constrained and 
$A\rightarrow A+B$ in (\ref{ttt}).
The solution associated to the point $\mathcal{C}$  instead is physical only if 
 \begin{equation} \label{}
\rho_0=0\, , \qquad B=(3-A)H_0^{2-\alpha}\,.
\end{equation}
It is interesting to note here that, since in the fixed points $\mathcal{A}$ and $\mathcal{B}$ the parameter $B$ is zero, these points represent states for the cosmology that indistinguishable from standard general relativity (GR). This, as we will see, will be an interesting feature in the physical interpretation of the orbits. 
\begin{table}
\caption{Coordinates of the fixed points, eigenvalues, and
solutions for the system (\ref{sistema non simpl1}-\ref{sistema non simpl3}).} \label{tavola punti fissi}
\begin{center}
\begin{tabular}{cccc}
\hline
\textbf{Point} & \textbf{Coordinates} $(x,y)$ & \textbf{Eigenvalues} \\ 
\hline
$\mathcal{A}$ &$\left[1, 0\right]$& $\displaystyle{\left\{
\frac{1}{2}\left( A-3 \right) \,\left( \alpha-2 \right) \,\left( 1
+ w \right) ,1 +3\,w  - A\,\left( 1 + w \right) \right\}}$\\
$\mathcal{B}$  &$\left[\frac{2\,A}{\left( 3 - A \right) \,\left( 1
+ 3\,w  \right)},0\right]$& $\{ \alpha (1+3  w) -2 A (w +1)+\mathds{A},   \alpha (1+3  w) -2 A (w +1)-\mathds{A},\}$\\
$\mathcal{C}$ & [0,1] & $\left\{
-2,\displaystyle{\frac{1}{2}\left( 3 - A \right) \,\left( \alpha
-2  \right) \,\left( 1 + w  \right)}  \right\}$\\
\hline\\[-4.mm]
\multicolumn{3}{c}{ $\mathds{A}=\sqrt{(\alpha (1+3  w) -2 A (w +1) )^2-4 (3 w+1)^2 (\alpha
   -2) ((A-3) w+A-1)}$}
\\[2mm]
\end{tabular}
\begin{tabular}{cccccc}
\hline
\textbf{Point} & \textbf{Scale Factor} & \textbf{Energy Density} &$G(t)$ & $\Lambda(t)$ \\ 
\hline\\[-1mm]
\multirow{2}{*}{$\mathcal{A}$} & $a= {a_0}\,{\left(t-{t_0} \right)}^{\beta }$ & 
\multirow{2}{*}{$\rho=\rho _0\left(t-t_0\right)^{\frac{6}{A-3}}$} & 
\multirow{2}{*}{${G=\frac{\left(t-t_0\right){}^{\frac{2 A}{3-A}}}{2(A-3) \pi \rho_0 (1+w)^2}}$} &
\multirow{2}{*}{$\displaystyle{\Lambda=\frac{4A (1+w)^{-2}}{(A-3)^2 (t-t_0)^2 }}$} \\
&$\beta  = \frac{2}{(3-A) \,\left( 1 + w  \right)}$& &\\
\\
\vspace{2mm}$\mathcal{B}$ & $a=a_{0}(t-t_{0})$&$\rho=\rho_0 (t-t_0)^{-3-3w} $ &
$\displaystyle{G=\frac{A \left(t-t_0\right){}^{3 w +1} }{4 \pi  \rho _0 (1+3w)}}$ & $\displaystyle{\Lambda=\frac{A}{(t-t_0)^2}}$\\
\\
\vspace{1mm}\multirow{2}{*}{$\mathcal{C}$} & $a=a_{0}\exp\left[H_0 (t-t_{0})\right]$& 
\multirow{2}{*}{$\rho=0 $} &
\multirow{2}{*}{$G=G_0$} & 
\multirow{2}{*}{$\Lambda=3 H_0^2$} \\
&$(A-3) H_0^2+B H_0^{\alpha }=0$&&\\[1mm]
\hline
\end{tabular}
\end{center}
\end{table}
The stability of the fixed points can be inferred with the
standard techniques of the dynamical system analysis and are
summarized in Table \ref{tbl:stabilit punti fissi1}. 

A global picture of the phase space that summarizes 
the results above can be found in Figures \ref{PS1} and \ref{fig2}.

\begin{table}[b]
\caption{Stability of the fixed points for the system 
(\ref{sistema non simpl1}-\ref{sistema non simpl3}). When the stability
depends on the barotropic factor both the characters of the fixed
points are shown. See the text for more details.} \centering
\bigskip
\begin{tabular}{c|ccc|cccccc}
\hline
 \multirow{2}{*}{Point}& & $\alpha<2$ &&&  & $\alpha>2$ &  \\
 & $A<\frac{1+3w}{1+w} $& $\frac{1+3w}{1+w}<A<3$ &$A>3$  &&  $A<\frac{1+3w}{1+w} $& $\frac{1+3w}{1+w}<A<3$ & $A>3$\\
\hline
 $\mathcal{A}$ & repeller & saddle& attractor&&  saddle & attractor & saddle \\
 $\mathcal{B}$  & saddle & repeller & repeller&&   attractor & saddle & saddle \\
 $\mathcal{C}$ & attractor & attractor & saddle&&   saddle& saddle& attractor \\ 
\hline
\end{tabular}\label{tbl:stabilit punti fissi1}
\end{table}

There  are also other global properties of the phase space that
we can deduce from equation (\ref{soluzioni}). 
Specifically the deceleration parameter $q$ can be written in
terms of the dynamical variables as
\begin{equation}\label{deceleration par}
q=-1-\frac{\dot{H}}{H^{2}}=-\frac{1 }{6}\left[2\,A + 2\,(A-3)
\,y +(A-3) \,x\,\left( 1 + 3\,w \right)\right]\;.
\end{equation}
This means that the line
\begin{equation}\label{linea accelarazione}
y = \frac{A}{ A-3} + \frac{1 + 3\,w   }{2}\,x,
\end{equation}
divides the phase plane in two regions
\begin{equation}
  A>3 \qquad\left\{%
\begin{array}{ll}
    y> \frac{A}{ A-3} + \frac{1 + 3\,w   }{2}\,x, \qquad (q<0),\\
     y< \frac{A}{ A-3} + \frac{1 + 3\,w   }{2}\,x,\qquad (q>0),\\
\end{array}%
\right. \end{equation}
\begin{equation}
A<3\qquad\left\{%
\begin{array}{ll}
    y> \frac{A}{ A-3} + \frac{1 + 3\,w   }{2}\,x, \qquad (q>0),\\
     y< \frac{A}{ A-3} + \frac{1 + 3\,w   }{2}\,x, \qquad (q<0),\\
\end{array}%
\right.
\end{equation}
in which  the decelerating factor is only positive or only
negative. This allows us to understand if a specific orbit includes
the transition between a decelerating an accelerating phase
typical of the dark energy. In addition, substituting the
coordinates of the fixed points, we can check if they represent a
decelerating or accelerating solution.  For point $\mathcal{C}$
the deceleration factor is always negative as expected by the
nature of the associated solution. Point $\mathcal{B}$ instead
lies always on the line (\ref{linea accelarazione}) and this is
again expected by the associated solution. Finally, point
$\mathcal{A}$ represents an accelerated expansion if
$\displaystyle{\frac{1+3w}{1+w}<A<3}$, a decelerated
expansion for $\displaystyle{A<\frac{1+3w}{1+w}}$ and a
accelerated contraction if $A>3$.
\begin{figure}[ht]
\centering
\subfigure[Case $A<1$.]{
\includegraphics[scale=0.38]{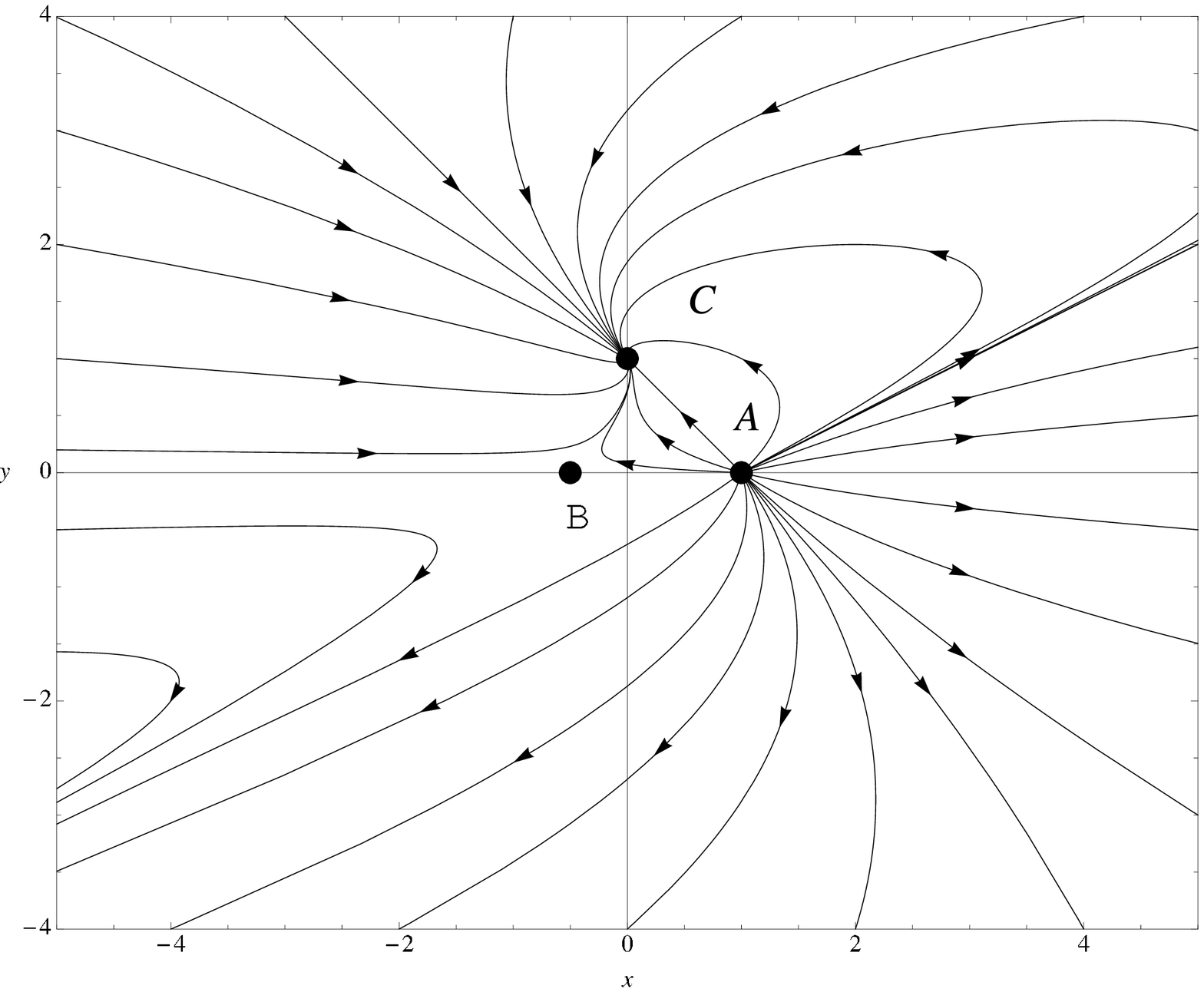}
\label{fig:subfig1}
}
\subfigure[Case $1<A<3$.]{
\includegraphics[scale=0.35]{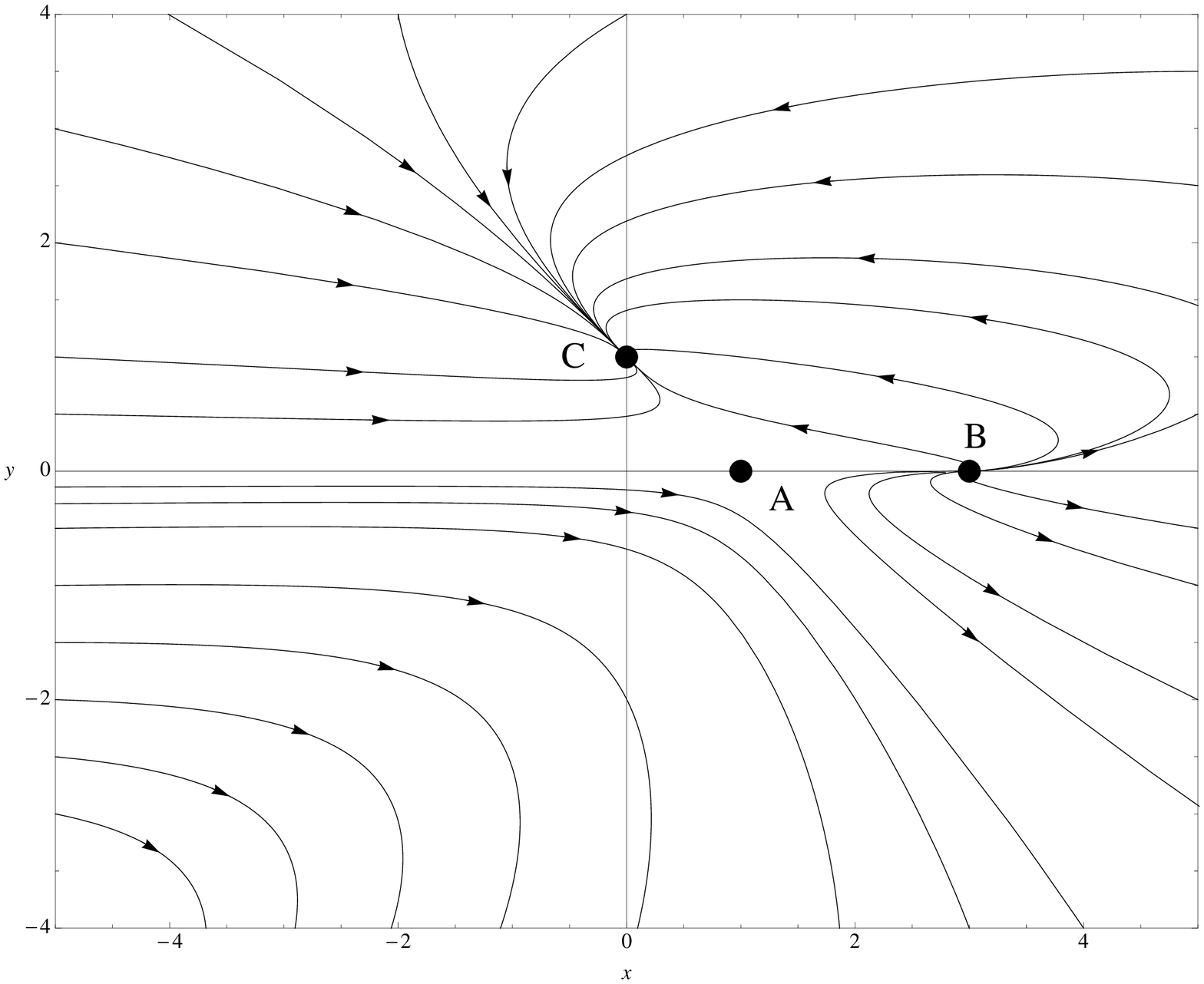}
\label{fig:subfig2}
}
\subfigure[Case $A>3$.]{
\includegraphics[scale=0.37]{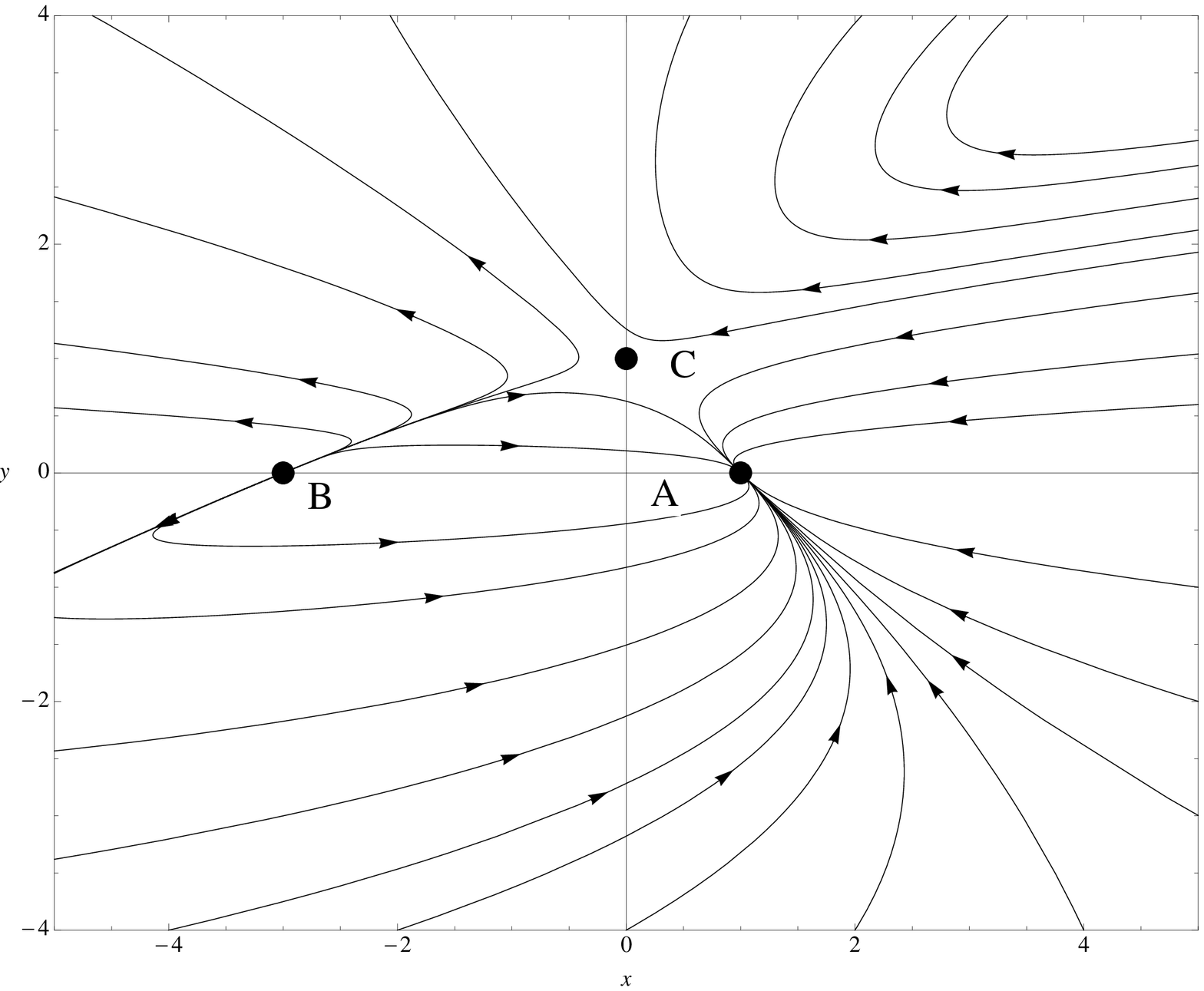}
\label{fig:subfig3}
}
\caption[Phase plots of the system  (\ref{sistema non simpl1}-\ref{sistema non simpl3}) for $\alpha<2$ and dust ($w=0$).]
{Phase plots of the system  (\ref{sistema non simpl1}-\ref{sistema non simpl3}) for $\alpha<2$ and dust ($w=0$).}\label{PS1}
\end{figure}

\begin{figure}[ht]
\centering
\subfigure[Case $A<1$.]{
\includegraphics[scale=0.34]{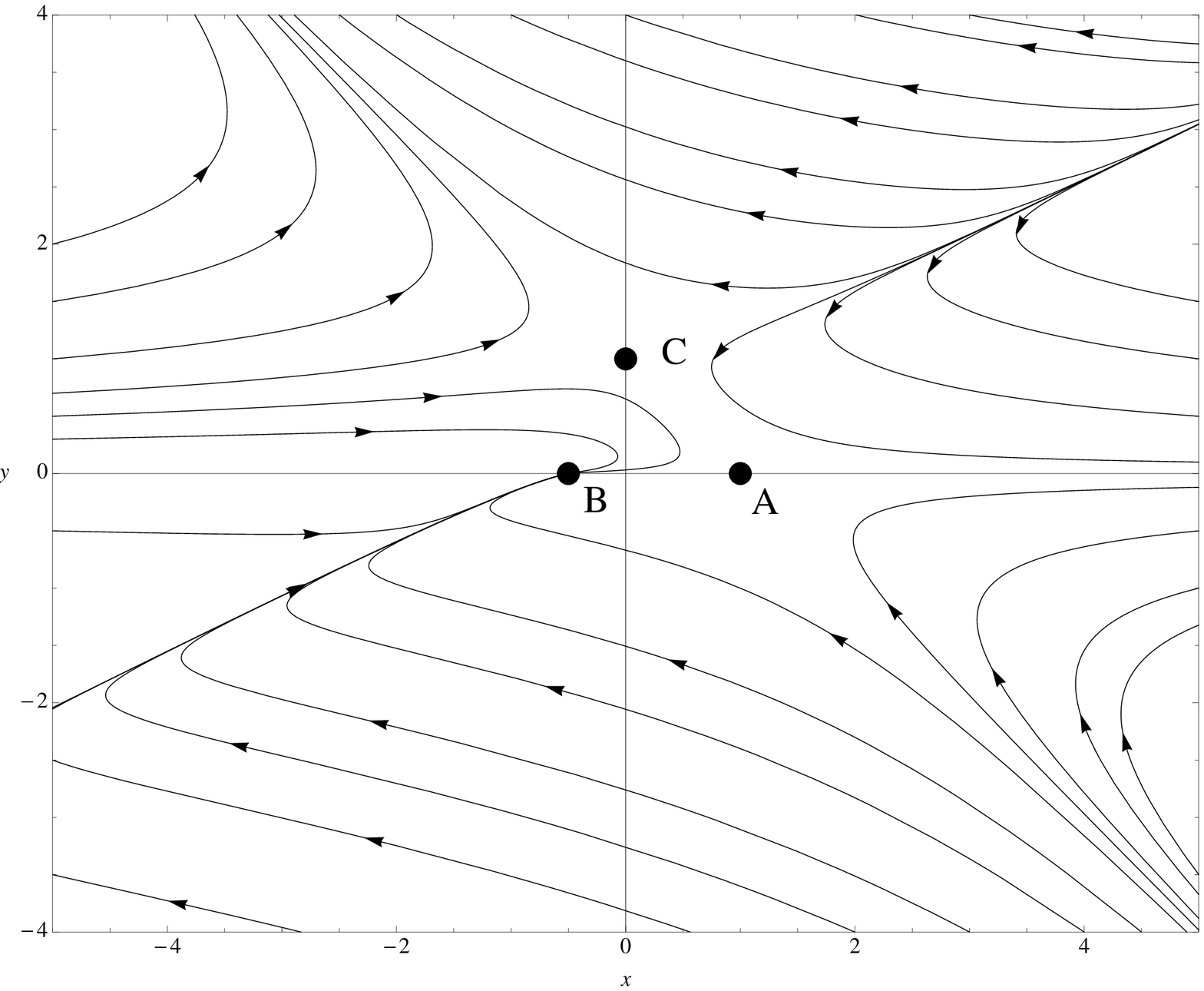}
\label{fig:subfig3}
}
\subfigure[Case $1<A<3$.]{
\includegraphics[scale=0.38]{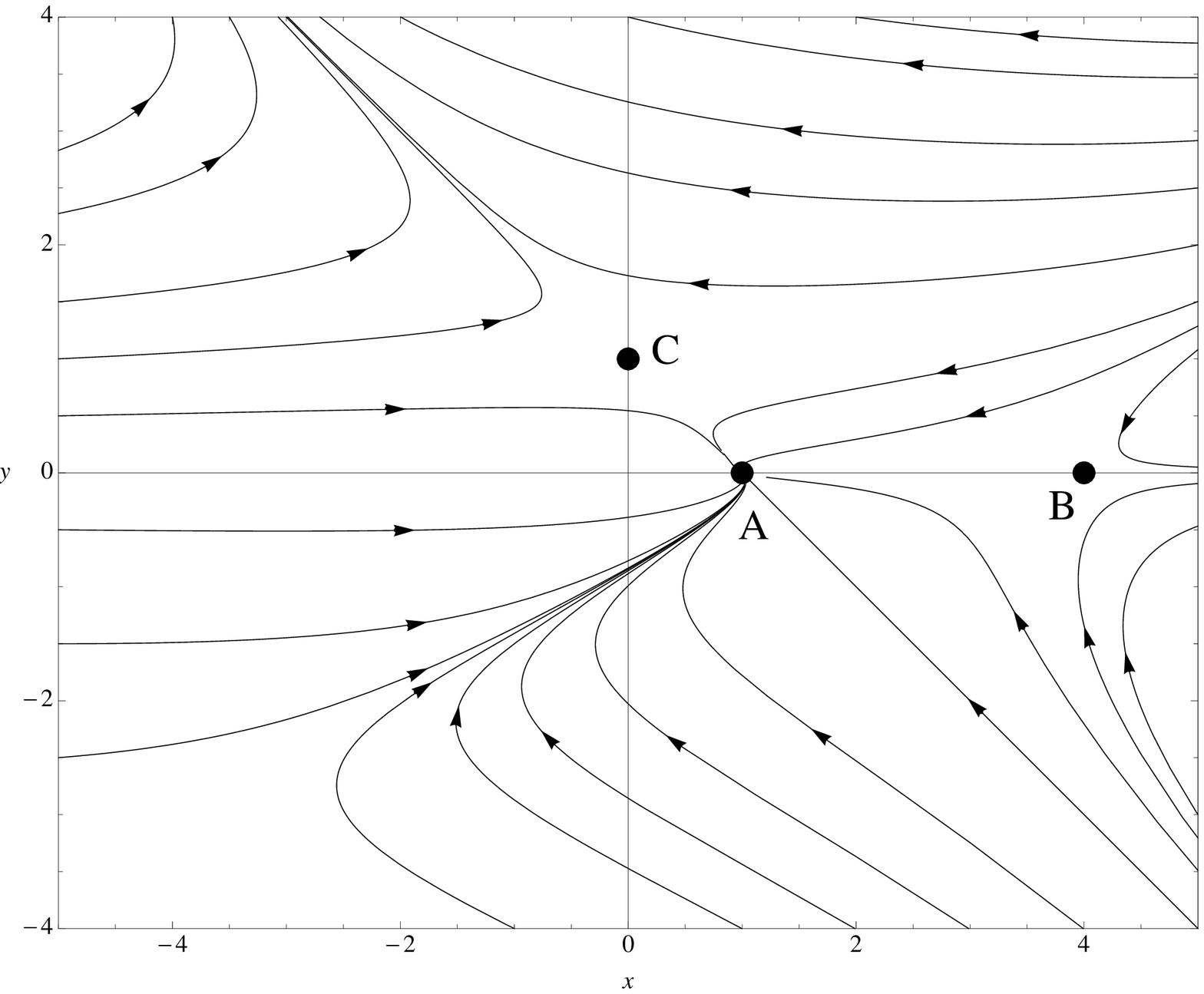}
\label{fig:subfig3}
}
\subfigure[Case $A>3$.]{
\includegraphics[scale=0.35]{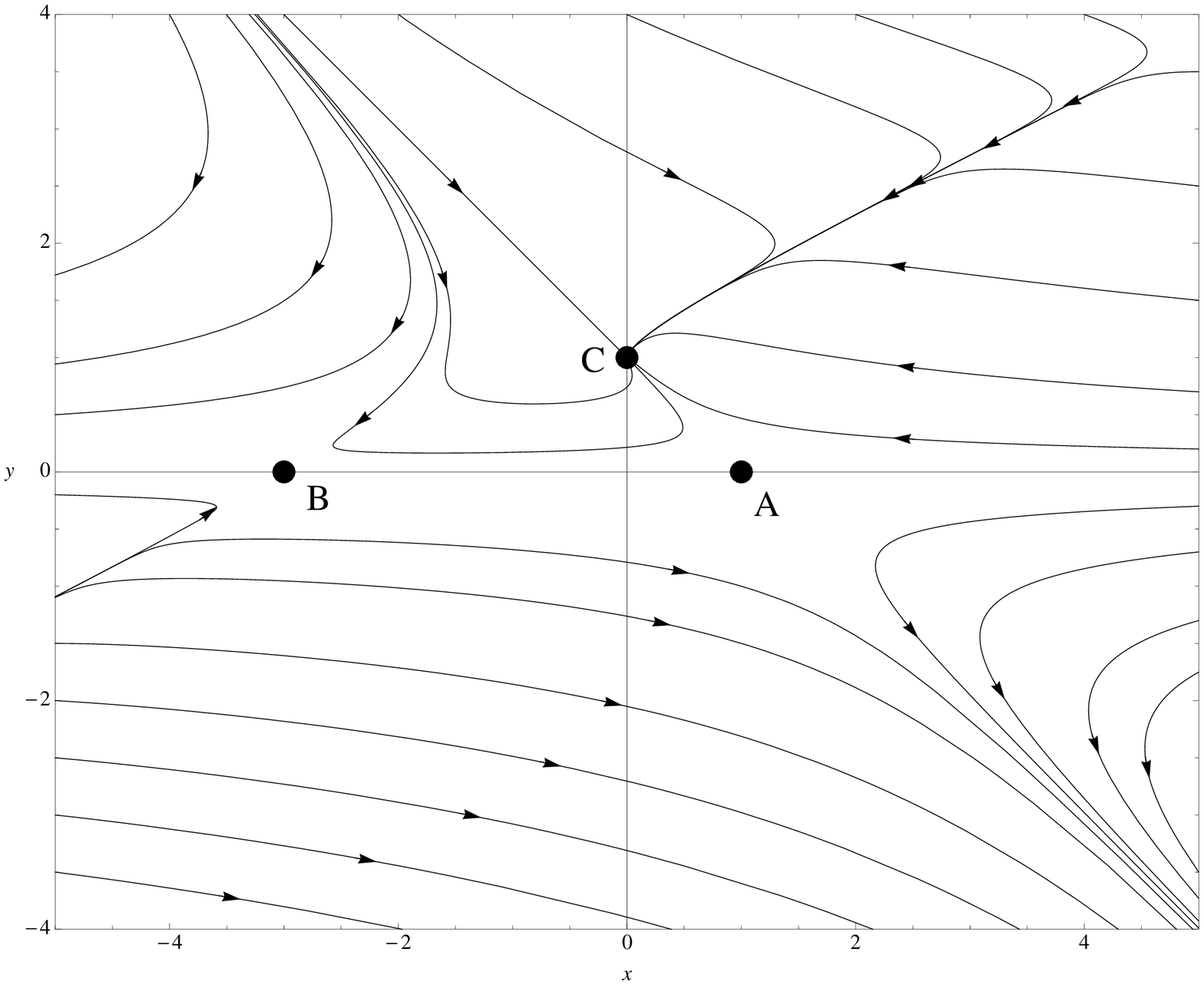}
\label{fig:subfig3}
}
\caption[Phase plots of the system  (\ref{sistema non simpl1}-\ref{sistema non simpl3}) for $\alpha>2$ and dust ($w=0$).]{Phase plots of the system  (\ref{sistema non simpl1}-\ref{sistema non simpl3}) for $\alpha>2$ and dust ($w=0$).}\label{fig2}
\end{figure}
\subsection{Asymptotic Analysis}
Looking at (\ref{sistema simpl3}) it appears clear that  the phase plane is not compact and it is
possible that the dynamical system (\ref{sistema simpl1}-\ref{sistema simpl2}) has a nontrivial asymptotic
structure. Thus the above discussion would be incomplete without
checking the existence of fixed points  at infinity (i.e. when the
variables $x$ or $y$ diverge) and calculating their stability.
Such points represent regimes in which one or more of the terms in
the Friedmann equation (\ref{useful fried}) become dominant and
should not be confused with the time asymptotics i.e.
$t\rightarrow\infty$.

The asymptotic analysis can be easily performed by {\it
compactifying} the phase space using the so called
Poincar$\acute{e}$ method \cite{Lefschetz77}.
The compactification can be achieved by transforming to polar
coordinates $(r(\cal{N}), \theta (\cal{N}))$:
\begin{equation}\label{coordinate compattificazione}
x\rightarrow r \cos\theta\;,\qquad \qquad y\rightarrow r
\sin\theta\;,
\end{equation}
and substituting $r\rightarrow\frac{{\cal R}}{1-{\cal R}}$ so that
the regime $r\rightarrow\infty$ corresponds to ${\cal
R}\rightarrow 1$. Using the coordinates (\ref{coordinate
compattificazione}) and taking the limit ${\cal R}\rightarrow 1$,
the equations (\ref{sistema simpl}) can be written as
\begin{eqnarray}\label{sistema asintotico}
 \mathcal{R} ' &=& \frac{1}{12} (A-3) (\alpha  \sin 2 \phi +\alpha  \cos 2 \phi -\alpha +4) (3 w  \cos \phi -2 \sin \phi +\cos \phi ),\label{rho'}\\
 \theta ' &=&  -\frac{ \alpha(A-3)}{4(\rho-1)}\,
        \left[\left( 1 +
 3\,w \right)\,\cos \theta - 2\,\sin \theta \right]\,
        \left( -1 + \cos 2\,\theta  - \sin 2\,\theta  \right)\label{theta'}.
\end{eqnarray}
It can be proven that the existence and the stability of the 
fixed points can be derived analyzing the  fixed point of the equation for $\theta$ 
\cite{Lefschetz77}. 
Setting $\theta'=0$, we obtain six fixed points
which are given in Table \ref{tavola punti fissi asintotici vuoto 1}.
The solution associated with the asymptotic fixed points can be
derived in the same way of the ones for the finite case and are also shown 
in Table \ref{tavola punti fissi asintotici vuoto 1}. For details on this derivation we refer the reader to 
the detailed discussion in \cite{Carloni:2004kp}.
The solutions associated to the asymptotic fixed points  are of two basic types. 
A first one is an exponential growth i.e. a deSitter phase (points $\mathcal{A}$-$\mathcal{B}$) and 
a second one whose growth and  decay depends on the values of $\alpha$ and $w$. 
In particular  they can represent bounces or cosmologies  in which the deceleration parameter changes sign.
\begin{table}
\caption{Asymptotic fixed points, $\theta$ coordinates and
solutions for the system (\ref{sistema non simpl1}-\ref{sistema non simpl3}).} \centering
\bigskip
\begin{tabular}{ccl}
\hline
 Point&  $\theta$ & Behavior \\ 
\hline
$\mathcal{A}_{\infty}$ & 0& $a= a_0\exp[\frac{1}{4}{C_1^{2}}\,( t - {t_0}) ]$\\
$\mathcal{B}_{\infty}$ & $\pi$& $a= a_0\exp\left[\frac{1}{4}{C_1^{2}}\,( t - {t_0}) \right]$\\
$\mathcal{C}_{\infty}$&$3\pi/4$ & $ a= a_0\exp\left[-\left(\frac{(\alpha -1) (t-t_0)}{2-\alpha}\right)^{\frac{\alpha -2}{\alpha -1}}\right]$\\
$\mathcal{D}_{\infty}$ &$7\pi/4$ & $ a= a_0\exp\left[-\left(\frac{(\alpha -1) (t-t_0)}{2-\alpha }\right)^{\frac{\alpha -2}{\alpha -1}}\right]$\\
$\mathcal{E}_{\infty}$ &$\arctan\left(\frac{1+3w}{2}\right)$&  $ a_0\exp\left[\left(t-t_0\right)^{r}\left(\frac{-3 (2-\alpha ) w  (3 w +2)-3 \alpha +2}{3(\alpha -2) (w +1) (3 w -1)}\right)^{r}\right]$\\
$\mathcal{F}_{\infty}$ &$\arctan\left(\frac{1+3w}{2}\right)+\pi$& $ a_0\exp\left[\left(t-t_0\right)^{r}\left(\frac{-3 (2-\alpha ) w  (3 w +2)-3 \alpha +2}{3(\alpha -2) (w +1) (3 w -1)}\right)^{r}\right]$\\ 
\multicolumn{3}{c}{$r=1+\frac{4}{9 (\alpha -2) w ^2+6 (\alpha -2) w -3 \alpha  +2}$}\\ 
\hline
     \end{tabular}
\label{tavola punti fissi asintotici vuoto 1}
\end{table}

\begin{table}
\caption{Stability of the asymptotic fixed points for
the system (\ref{sistema non simpl1}-\ref{sistema non simpl3}).}
\centering
\bigskip
\begin{tabular}{c|ccc|ccc}
\hline
\multirow{2}{*}{Point} & &$A<3$ &&& $A>3$\\ 
& $\alpha<0$ & $0<\alpha<2$ & $\alpha>2$ & $\alpha<0$ & $0<\alpha<2$ & $\alpha>2$  \\ 
\hline
$\mathcal{A}_{\infty}$ & attractor& saddle & saddle & repeller & saddle & saddle\\
$\mathcal{B}_{\infty}$  &repeller& saddle & saddle & attractor & saddle & saddle\\
$\mathcal{C}_{\infty}$ & saddle & repeller & saddle & saddle & attractor & saddle\\
$\mathcal{D}_{\infty}$ & saddle& attractor & saddle & saddle  & repeller & saddle\\
\hline
\end{tabular}
\label{tavola stasbilita punti fissi asintotici 1}
\end{table}
Finally, the stability of the asymptotic fixed points can be obtained with
the standard methods of the dynamical system. The results for the
first four fixed points
($\mathcal{A}_{\infty}$-$\mathcal{D}_{\infty}$) is given in Table
\ref{tavola stasbilita punti fissi asintotici 1}. 

The same is not true for the last two fixed points
($\mathcal{E}_{\infty}$, $\mathcal{F}_{\infty}$) whose stability
do depend on $w$. The character  of these points is
complicated by the fact that the value of $\mathcal{R}'$, whose
sign is connected to the stability in the radial direction is
zero. Fortunately this problem can be solved noting that the next
to leading term in the full the equation for $\mathcal{R}'$ is
finite and different from zero and can give information on the
behavior of the function $\mathcal{R}'$ nearby $\mathcal{R}=1$. 
The stability thus obtained is shown in  Tables \ref{tavola stasbilita punti fissi asintotici EF 1} and \ref{tavola stasbilita punti fissi asintotici EF 2}.
\begin{table}
\caption{Stability of the asymptotic fixed point
$\mathcal{E}_{\infty}$.} \centering
\bigskip
\begin{tabular}{llllllll}
\hline
&  attractor & repeller & saddle\\   
\hline
&$\alpha \leq -10\land 3<A<\frac{3 \alpha -6}{\alpha +10}$ & &\\
$w= 0$ &$-10<\alpha <0\land  A>3$& $-10<\alpha<0\land A<\frac{3 \alpha -6}{\alpha +10}$ & otherwise\\
 & $\alpha>0 \land \frac{3 \alpha -6}{\alpha +10}<A<3$&  & \\ 
\multirow{2}{*}{$w= 1/3$}& $\alpha<0 \land A >3$&  \multirow{2}{*}{$\alpha<0 \land A<0$}& \multirow{2}{*}{otherwise}\\
&  $\alpha>0 \land 0<A <3$& & \\ 
& $\alpha <0 \land A>3$&& \\ 
$w= 1$& $0<\alpha \leq 5 \land \frac{3 \alpha -6}{\alpha -5}<A<3$& \multirow{1}{*}{$\begin{array}{c} \alpha <0\land A<\frac{3 \alpha -6}{\alpha -5}\\ \alpha >5\land A>\frac{3 \alpha -6}{\alpha -5}\end{array}$}& otherwise\\
&  $\alpha> 5\land A<3$&& \\
\hline
\end{tabular}
\label{tavola stasbilita punti fissi asintotici EF 1}
\end{table}

\begin{table}
\caption{Stability of the asymptotic fixed point
$\mathcal{F}_{\infty}$.} \centering
\bigskip
\begin{tabular}{cccccc}
\hline
&  attractor & repeller & saddle\\   
\hline
\multirow{3}{*}{$w= 0$} & $ \alpha <-10\land A<3$& 
\multirow{3}{*}{$\begin{array}{c} \alpha <-10\land A>\frac{3 \alpha -6}{\alpha +10} \\
   \alpha >0\land A<\frac{3 \alpha -6}{\alpha +10}\\
\end{array}
$} &
\multirow{3}{*}{ otherwise} \\ 
  & $-10< \alpha <0\land \frac{3 \alpha -6}{\alpha +10}<A<3$ & & \\ 
      & $\alpha>0 \land A>3$& & \\  
\multirow{2}{*}{$w= 1/3$}& $\alpha<0 \land 0<A<3$& \multirow{2}{*}{$\alpha <0\land A<0$}& \multirow{2}{*}{otherwise}\\
&  $\alpha>0\land A>3$& & \\ 
\multirow{3}{*}{$w= 1$} &$\alpha<0 \land \frac{3 \alpha -6}{\alpha -5}<A<3$ &\multirow{3}{*}{$0<\alpha<5\land A<\frac{3 \alpha -6}{\alpha -5}$}  & \multirow{3}{*}{otherwise} \\ 
   &  $0<\alpha\leq5\land A>3$& & \\
   &  $\alpha>5\land 3< A<3\frac{3 \alpha -6}{\alpha -5}$& & \\
\hline
\end{tabular}
\label{tavola stasbilita punti fissi asintotici EF 2}
\end{table}
Now that the asymptotic fixed points and their stability has been determined let us look in more detail to their physical interpretation. 

As said before, these points are characterized by one or more variables to become infinite. In terms of the definitions (\ref{DynVar}), this corresponds to the fact that either the quantities in the numerators are infinite or the ones in the denominators approach zero. In the first case we are probably seeing some kind of singularity of the model. In the second we are seeing a change in sign of the expansion, which in turns corresponds to a maximum or a minimum of the scale factor. It is important to bear in mind that, because of our definition of the time variable,  when an orbits ``reaches'' an infinite point the time coordinate changes sign and the Universe follows the orbit with a reversed orientation. In some sense one can picture this transition as the fact that the asymptotic points are the doorway to a mirror phase space in which the orbit orientation and stability of the fixed points are reversed.

In the pure GR framework, where the cosmological equation predict  (almost) always a monotonic behavior of the scale factor, extrema of the scale factors can occur only at the origin of time. We usually talk then of ``bounces'' or ``(re)collapsing universes'' depending on the sing of the second derivative of the scale factor (or $\dot{H}$). In more complex cosmological models, like e.g. $f(R)$ gravity \cite{Carloni:2005ii}, the behavior is more complicated and the scale factor can have in principle a series of extrema located at a generic instant. This is the case also for the RG cosmologies. However, we will retain the traditional names to indicate such features of the scale factor.

Let us focus, for example, on the conditions to have a bounce i.e. a situation in which $\dot{a}=0$ and $\ddot{a}>0$ . In terms of the dynamical variables these can be translated in the requirement that one has an asymptotic attractor characterized by $H=0=\dot{a}$ and that this attractor lays in the part of the phase space for which $-(1+q)<0$  which is determined via (\ref{linea accelarazione}). Using this condition it turns out, for example, that for $\alpha>0$ only $\mathcal{D}_{\infty}$ can represent a bounce. The results for the other points can be found in Table \ref{AsyCosm}.

\begin{table}
\caption{Physical interpretation of the Asymptotic Attractors.}
\centering
\bigskip
\begin{tabular}{c|cccccc}
\hline
Point & Bounce &(Re)Collapsing& Singularity\\ 
\hline
$\mathcal{A}_{\infty}$ & No & No & Yes \\
 & &  & \\
$\mathcal{B}_{\infty}$ & No & No & Yes \\
 & &  & \\
$\mathcal{C}_{\infty}$ & No &  $0<\alpha<2$, $A>3$ & Otherwise\\
 & &  & \\
$\mathcal{D}_{\infty}$ & $0<\alpha<2$, $A>3$&No & Otherwise \\
 & &  & \\
\multirow{2}{*}{$\mathcal{E}_{\infty}  (w=0)$ } & $\alpha \leq -10\land 3<A<\frac{3 \alpha -6}{\alpha +10}$  
& $0<\alpha <\frac{6+10A}{3-A}\land  -\frac{3}{5}<A<0$  & Otherwise\\
 &$-10<\alpha <0\land  A>3$ &$0 < A < 3$, $0 <\alpha< 2$ & \\
  & &  & \\
\multirow{2}{*}{$\mathcal{F}_{\infty} (w=0)$} & $\alpha \leq -10\land 3<A<\frac{3 \alpha -6}{\alpha +10}$ 
& $0<\alpha <\frac{6+10A}{3-A}\land  -\frac{3}{5}<A<0$   & Otherwise \\
 &$-10<\alpha <0\land  A>3$ &$0 < A < 3$, $0 <\alpha< 2$   & \\
  & &  & \\
$\mathcal{E}_{\infty}  (w=1/3)$  & $\alpha<0 \land A>3$  & $0<\alpha<2 \land 0<A<3$  & Otherwise\\
  & &  & \\
$\mathcal{F}_{\infty} (w=1/3)$ & $\alpha<0 \land A>3$  & $0<\alpha<2 \land 0<A<3$ & Otherwise \\
 & &  & \\
\multirow{2}{*}{$\mathcal{E}_{\infty}  (w=1)$} & $\alpha \leq -10\land 3<A<\frac{3 \alpha -6}{\alpha +10}$  
& $0<\alpha <\frac{6+10A}{3-A}\land  -\frac{3}{5}<A<0$  & Otherwise\\
 &$-10<\alpha <0\land  A>3$ &$0 < A < 3$, $0 <\alpha< 2$ & \\
  & &  & \\
\multirow{2}{*}{$\mathcal{F}_{\infty} (w=1)$} & $\alpha \leq -10\land 3<A<\frac{3 \alpha -6}{\alpha +10}$ 
& $0<\alpha <\frac{6+10A}{3-A}\land  -\frac{3}{5}<A<0$   & Otherwise \\
 &$-10<\alpha <0\land  A>3$ &$0 < A < 3$, $0 <\alpha< 2$   & \\\hline
\end{tabular}
\label{AsyCosm}
\end{table}
\begin{figure}[ht]
\centering
\includegraphics[scale=0.43]{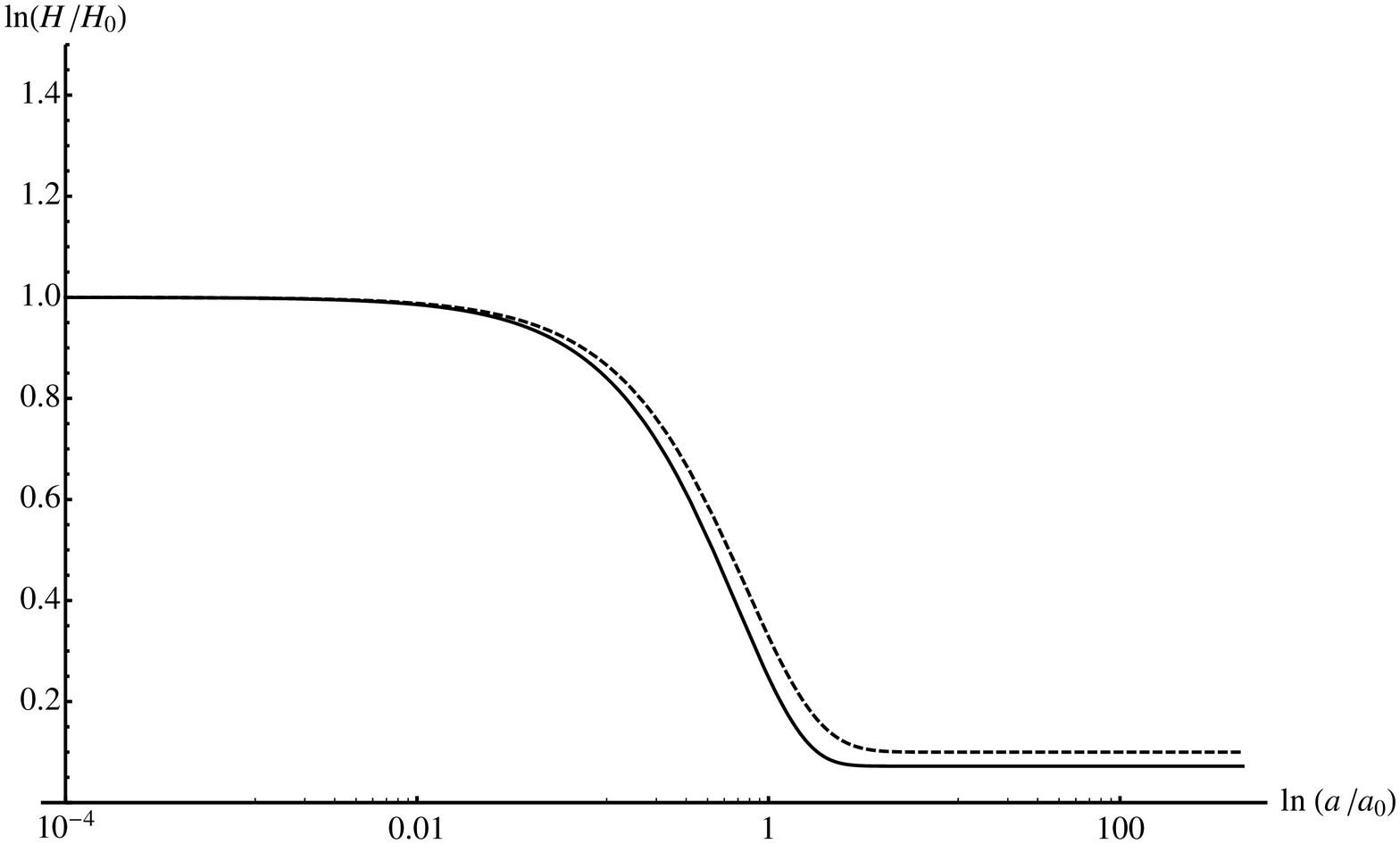}\\
\includegraphics[scale=0.43]{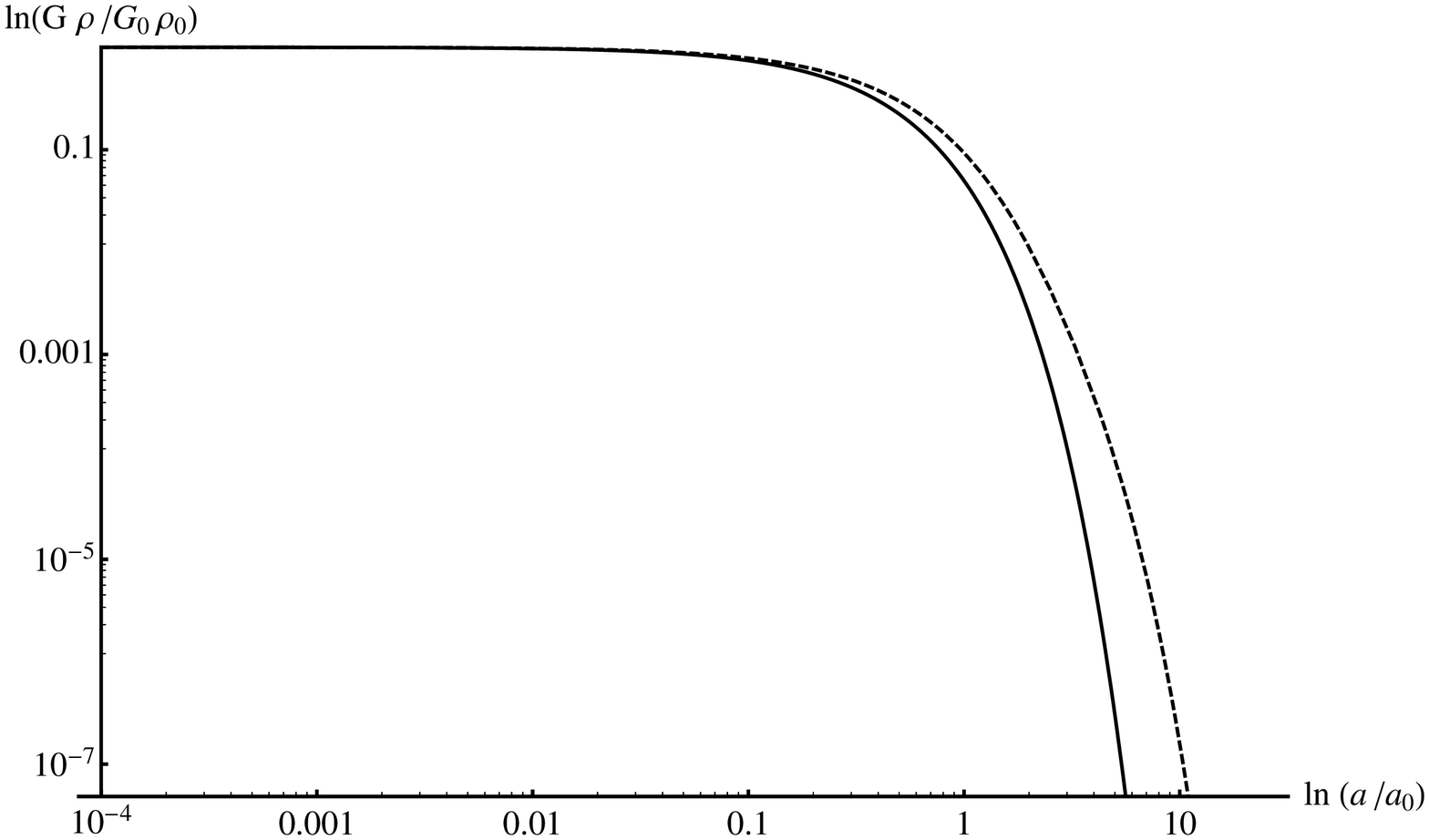}
\caption{Evolution of  some key cosmological quantities in RG cosmologies with $w=0$ $A=1/2$, $\alpha=1$ (dashed) compared with the ones of GR$\Lambda$ (solid). We have chosen as initial condition the phase space point $(x=1.1,y=0.1)$ present in both the phase spaces. The index 0 is associated to the value of all the quantities in this point. Upper panel: Semi logarithmic plot of $H/H_0$ as function of $\ln (a/a_0)$. Lower panel: Logarithmic plot of $ \frac{8}{3}\pi G \rho$ as a function of $\ln (a/a_0)$. The energy density is not plotted because its behavior is exactly matched in the two models.}
\label{PlotCosm1}
\end{figure}

\begin{figure}[ht]
\centering
\includegraphics[scale=0.5]{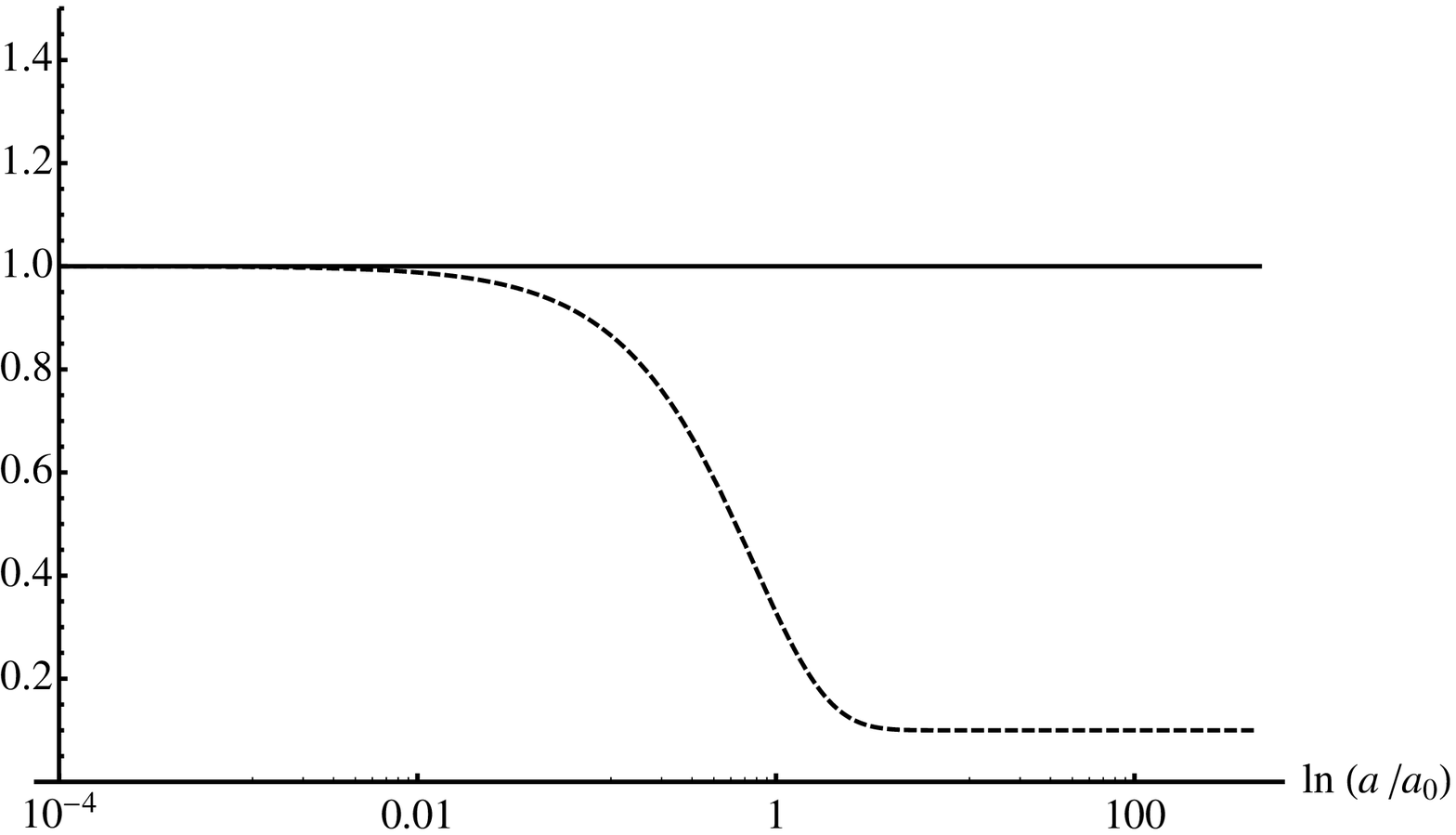}
\caption
{Evolution of the term $\frac{BH^{\alpha}}{3-A}$  in RG cosmologies with $w=0$ $A=1/2$, $\alpha=1$, $B=2$ (dashed) compared with the $\Lambda$  term in GR (solid). We have chosen as initial condition the phase space point $(x=1.1,y=0.1)$ and  the index 0 is associated to the value of all the quantities in this point. 
}
\label{PlotCosm3}
\end{figure}
%
\section{Conclusions}
In this paper we have investigated the effect on cosmological scale of a quantum gravity related decaying of the cosmological constant.  
This investigation was performed  using the Dynamical System Approach that allows a semi-quantitative treatment of the model via the construction of a phase space directly connected to the behavior of the equations. 
The phase space is two dimensional and it contains a total of nine fixed point, of which three are finite and six are asymptotical. 
The stability of these points is found to depend on the parameters $A$ and $\alpha$, but not on $B$ which has a marginal position in the entire analysis. 
The determination of the fixed points and their stability allows us to infer the shape of the phase space orbits and to deduce the qualitative features of the cosmic histories possible in this models.

Among these, one is particularly relevant in terms of the problem of  dark energy 
domination i.e. the presence of  a cosmic history characterized by a first phase in which the scale factor grows as a power law with exponent included in $]0,1[$ followed by a second phase with a faster growth. In terms of the  features of the phase space that  such scenario can be realized if one has an unstable fixed point representing the first phase and an attractor representing the second\footnote{The second point could  be unstable as well, but then the set of the initial condition able to realize such scenario would be smaller and more difficult to calculate.}. Looking at  Tables \ref{tbl:stabilit punti fissi1} - \ref{tavola stasbilita punti fissi asintotici EF 2} 
one can easily see that there is only is one set of the parameters $(A, \alpha)$ for which this can be realized: $(A<\frac{1+3w}{1+w}, \; \alpha<2)$, for which $\mathcal{A}$ is a repeller.
It is reassuring to notice that in this case  $A$ is of the order  unity  and, as 
$A=\xi^2\lambda_\ast$, we deduce that also $\xi$ is  of the order unity being $\lambda_\ast < 1/2$. 

When these conditions are satisfied there can exist, depending on initial conditions, one orbit that starts at $\mathcal{A}$ and  either go directly to $\mathcal{C}$ or bounce  at $\mathcal{E}_{\infty}$. The first case can be seen as a ``classical'' Friedmann-de Sitter transition. In the second case the cosmological evolution could be richer because  the solution  $\mathcal{E}_{\infty}$ can indicate, depending on the sign of $\alpha$, a growth whose rate saturates or an expansion phase comparable with a de Sitter one. One could interpret this as a ''two regimes`` dark energy phase.  Note that, as we have seen, in the fixed points $\mathcal{A}$ and $\mathcal{B}$ the parameter $B$ is zero these points represent states for the cosmology that are indistinguishable from standard general relativity. This means that our model in the neighborhood of these points is indistinguishable from the standard cosmology. 
 An example of the evolution of some key cosmological quantities compared with the standard de Sitter cosmology is given in Figure \ref{PlotCosm1}, and in Figure \ref{PlotCosm3} we compare the additional RG term in (\ref{useful fried}) with the standard cosmological constant.
%

{It is important to emphasize that the standard experimental value of Newton's constant,
$G_{\rm exp}$, does not coincide with the value $G(k=\xi H_0)$ which is relevant for
cosmology today. $G_{\rm exp}$ is measured (today) at 
$k_{\rm exp}\propto \ell^{-1}$ where the length $\ell \equiv \ell_{\rm sol}$ is a 
typical solar system length scale, say $10^{12}$ m. Thus, in terms of the running Newton constant,
$G_{\rm exp} = G(k=\xi'/\ell_{\rm sol})$,
since $\ell_{\rm sol}\ll H_0$.
It is only the cosmological quantity $G(k=\xi H)$ which dynamically evolves in the fixed point regime,
not $G_{\rm exp}$.  
This remark entails that the dynamical evolution of the cosmological Newton constant in the 
recent past does not ruin the predictions about primordial nucleosynthesis which requires that
$G(k=\xi H_{\rm nucl})$ coincides with $G_{\rm exp}$ rather precisely. 
In fact, at the time $t=t_{\rm nucl}$ 
of nucleosynthesis 
for which $H=H(t=t_{\rm nucl})$,
the cosmological Newton 
constant was indeed $G(k=\xi H_{\rm nucl})\approx G_{\rm exp}$ 
since $ct_{\rm nucl} \approx H_{\rm nucl}^{-1}$ and $\ell_{\rm sol}$
are of the same order of magnitude.}

The phase space structure also sheds light on aspects of the type of RG cosmologies not directly related to the problem of dark energy. For example, it is clear that it is possible to interpret the transition to a dark era as the beginning of a primordial inflationary phase. 
{Our results then imply that there is only one set of values that allows a ``graceful exit'' i.e. the transition from inflation to a Friedmann cosmology. Specifically in the case $(A<\frac{1+3w}{1+w}, \; \alpha>2)$ we can have this kind of scenario. However, since there is only one point which represents accelerating expansion one can see that the RG model can be used to represent {\em either} the inflationary era {\em or} the dark energy era, but not both.}

Finally the analysis of the asymptotic fixed points gives information on the possibility of changes in the sign of expansion rates in this type of cosmologies. In general relativity such phases are normally associated  either to the so-called "bounces" and are associated to cyclic Universes (when $\dot{H}>0$) or to recollapsing universes (when $\dot{H}<0$). In modified theories of gravity changes in the sign of $H$ can occur also when the size of the Universe is large because the scale factor does not need to be monotonic \cite{Carloni:2005ii}. 
In terms of the phase space  this kind of behavior is associated to the presence of asymptotic attractors and the sign of the quantity $-(1+q)$  as expressed in (\ref{deceleration par}) in the fixed point. In particular we have that if  $-(1+q)>0$ we have a deceleration followed by an acceleration and if  $-(1+q)<0$ the opposite situation. A quick analysis shows, for example, that only some of the asymptotic attractors for very specific values of the parameters can give origin to a bounce.

The previous results point towards a cosmology with interesting features that, in our opinion, deserves some more study. In particular one could try to test the transition to Dark Age we have found against the SnIa data to obtain some more constraints on the free parameters. This, and other issues, will be discussed in a following work.

\bibliography{t2}
\end{document}